\def\beq{\begin{equation}}
\def\ee{\end{equation}}
\def\ba{\begin{eqnarray}}
\def\ea{\end{eqnarray}}
\def\figloc#1#2{\epsfysize=3in
    \centerline{\epsfbox{fig#1.ps}}
    \centerline{Figure #1}
    {\raggedright\it   #2 }
    \bigskip
    }

\documentstyle[preprint,aps,epsf]{revtex}
\begin{document}

\title{Varieties of Quantum Measurement}
\author{W. G. Unruh }
\address{
 CIAR Cosmology Program\\
Dept. of Physics\\
University of B. C.\\
Vancouver, Canada V6T 1Z1\\
email unruh@physics.ubc.ca}

\maketitle

 ~

 ~

\begin{abstract}
Quantum measurement theory has fallen under the resticting influence of the
attempt to explain the fundamental axioms of quantum theory in terms of
the theory itself. This has not only led to confusion but has also restricted
our attention to a limited class of measurements. This paper outlines some
of the novel types of measurements which fall outside the usual textbook
description.
\end{abstract}

The problem of quantum measurement has been with us since the foundations
of the the theory were laid in the mid 1920's. It has generated much
discussion,
with little resolution of the questions raised. I will argue in this talk
that this situation has arisen in part because of the confusion brought about
by giving
two very different concepts the same name, with the expected result that
the valid questions related to the two concepts become entangled. It
furthermore
has led to a restriction on the types of measurements considered within the
theory. In this talk I am not going to propose any radical or even very
new interpretations of the theory of quantum mechanics. I am rather going
to engage in an ancient philosophical past--time, name to propose that
we use distinct terms for distinct concepts. I am then going to review
some of the novel insights which have been obtained recently ( especially
by the group around Aharonov) regarding some novel types of measurement.

\section{Measurement, Determination and Knowledge}

The concept of measurement in quantum mechanics has had a long and confused
history.  There are essentially two separate concepts which have been
conflated under the same title, concepts with a very different status
in the  theory {\it a priori}. In part the intense confusion surrounding
the word results from the attempt to reconcile these two different concepts,
or rather to apply the properties of the one concept to the other.

The first concept subsumed under the term measurement is an axiomatic
concept. Quantum mechanics, as with all of our theories in physics, is
based on a set of mathematical structures. In the case of quantum mechanics,
these structures are those of complex Hilbert spaces, and operators on
those Hilbert spaces. In addition to such mathematical structures, the
theory must also make contact with the physical world. Structures in the
theory must be correlated by structures in our experience of the world
itself. As with all theories, quantum mechanics is a means of answering
questions about our experiences of the world.
 Furthermore they are questions which  are related
to the particulars and peculiarities of the actual world we live in.
The theory requires a mapping the mathematical
structure onto   our experiences. As in all physical theories this takes the
form both of a general
map, of general structures of the world which we expect to have
a broad range of validity, and structures which reflect the particulars
and peculiarities of our experiences.

In classical physics, the former is called the dynamical theory, while the
latter
is called the initial conditions. The theory encompasses the identification
of dynamic variables and equations of motion, while the "initial conditions"
encompass those aspects of our experience which are felt to be peculiar
to the individual time and place of those experiences.

Quantum mechanics contains both of these aspects as well, but in a very
different form from that of classical physics.  The dynamics is represented
by the operators, while, in the simplest case, the particulars of the
situation is represented by the vector in the Hilbert space, the wave function.
I will denote these particulars by the term knowledge or conditions, rather
than the term "initial conditions", since as we will see, conditions need
not be initial, nor are they in general equivalent to initial conditions
(as they are in classical physics).

In addition to explanations, the theory must produce answers, must give us the
answers to questions
that we may have about the physical situations that we are interested in.
It is here that the theory actually makes contact with the physical world.
In quantum mechanics these answers are in terms of probabilities. The
usual phrasology goes something like " When one measures a quantity, and
the system is in the state $|\phi>$, the outcome of that measurement is
one of the eigenvalues, say $a$,  of the operator, say {\bf A}, representing
the physical variable measured, and the probability is given by the usual
expression $|<a|\phi>|$."

However, the word "measure" brings with it the image of a physical process.
Measurements are performed by means of measuring apparatuses. As aspects
of the physical world, such measuring apparatuses should themselves be
describable
by quantum mechanics itself. But it is difficult to have a system in which
at the same time a concept is an axiomatic feature of the theory, and
one describable by the theory. I would therefore suggest that the word
"determine"
be used instead for this axiomatic feature of the theory. Thus I would
rephrase the above sentence as `` When one determines a quantity, and the
knowledge ( or conditions) under which one wishes to determine that quantity
are represented by the vector $|\phi>$, then the determination of a quantity
represented by {\bf A}  gives one of eigenvalues of {\bf A}, say $a$ with
probability $|<a|\phi>|^2$."

Determination, in this axiomatic sense, says nothing about how the
determination
was made. It is simply a statement of a mapping from the theory to our
experience, in which some knowledge sets the conditions on the questions
we wish to ask, and some knowledge represents the answers to the questions
we want to ask.

What then is a measurement? I will reserve the term measurement for a physical
process, a process which is describable in terms of quantum theory itself.
A measurement is a process in which one has two separate physical systems,
represented by two separate sets of dynamical operators. Furthermore the
dynamical evolution is such that, given certain conditions on the measuring
apparatus, a determination of some quantity associated with the measuring
apparatus will give information about the system of interest.

Von Neumann \cite{vonneumann} showed that under certain conditions, a
measurement on a system
could be treated as a determination of that system. I.e., certain types
of measurement ( in which one makes a determination of some aspect of the
measuring apparatus only) acted in all ways as though one had instead made
a determination of the system itself. There is a consistency in quantum
mechanics, such that the axiomatic concept I call determination, is closely
related to the physical process I call measurement. However, notice that
in von Neumann's analysis, one has not done away with the concept of
determination.
One still must apply the axiomatic concept of determination to the measuring
apparatus before one can draw any conclusions at all from the theory. It
is just that such measurements allow us to reduce a complicated system
( apparatus plus system of interest) to a simple system ( the system of
interest alone) under certain conditions.
This mapping of a complex system onto a simpler system does not however
in any way change the requirement for the axiomatic concept of
``determination". It simply changes the system to which we need to apply the
concept.

At least in part the measurement problem in quantum mechanics is the disquiet
that physicists feel for the concept of ``determination". It feels like
an extra and extraneous concept, a non-physical concept. In classical physics,
one can imagine that the theory and reality are in complete correspondence.
The position of a particle really is a number, and our experience of that
position is simply the experience of that number. The physical map from
experience to theory is just an identification of those numbers in the
theory with the experience. (That some fairly sophisticated manipulations
of experience are necessary to extract that number is a technical detail.)
In quantum mechanics on the other hand, there seems to be no direct map
from our experience to the theory. The operators themselves have far too
much structure for experience. The state, or Hilbert space vector itself,
has the wrong properties to map onto our experience. The only map is the
rather indirect and seemingly unnatural one of ``determination".
One would like either to subsume determination under some physical concept
of the theory ( but that would loose the only relation between experience
and the theory that the theory contains) or to introduce some other relation
between the theory and experience from which one could derive `determination'
in a natural way. That neither of these objectives has ever been achieved
is a large part of the `problem of measurement' in quantum mechanics.

However, I do not want to spend any more of my time on this issue. Rather
I want to point out the the concern about this problem has warped our thinking
about quantum mechanics and about the types of measurement possible in
the theory.
Because the von-Neumann type of measurement creates the possibility of
reduction of a complex system to a simpler system, the idea has become
implanted that all measurement must be of the same sort. Because determination
has a certain form, measurement must have the same form seems
to be the thinking. However it is
becoming clear, especially through the work of the group around Aharonov,
that this is too restrictive.

Measurement is a physical process by which one has two system interacting,
and by making a determination on the one system, one can obtain information
about
the other system. In certain cases, the information obtained is the same
as a determination, but in other cases it can differ significantly.
Furthermore,
because of the similarity of wave mechanics to classical wave theory, the
impression has also arisen that conditions in quantum mechanics are entirely
equivalent to conditions in classical mechanics, namely initial conditions.
Let me look at the last case first.

It has long been known to some ( but ignored or resisted by most) that
the conditions in quantum mechanics differ significantly from those of
classical physics\cite{ABL}. In classical physics, all conditions can, by use
of
the equation of motion, be mapped onto initial conditions. Whether one
measures the position now and the momentum two days hence, or measures
them both now is really irrelevant. For any condition, imposed at time,
one can always, by use of the equations of motion, produce initial conditions
which are entirely equivalent in all of their predictions to those general
conditions. However, as Aharonov, Bergmann, and Liebowitz\cite{ABL} already
showed
about 30 years ago ( and as has been independently rediscovered often
since---e.g.,
\cite{unruh-nyas})
setting conditions at different times may not be equivalent to any initial
conditions. The simplest example
is that of a spin 1/2 particle whose x component of spin is known at 9AM
and y component at 11 AM. Say both are known to have value $+1/2$. The
probability that if one determines the component $\cos(\theta) S_x
+\sin(\theta)S_y$
at 10AM, the answer will be $+1/2$ is
\beq
P_{S_\theta=1/2}= {(1+\cos(\theta)(1+\sin(\theta)\over
 2(1+\sin(\theta)\cos(\theta))} .
\ee
 Note there exist
no initial condition--- wave function or density matrix--- which would
give this answer. It is unity for both $\theta=0$ and for $\theta=\pi/2$.
The conclusion drawn from this simple example  is true in general-
 conditions in quantum mechanics
are not equivalent to initial conditions.

Already at the last NY Academy of Sciences meeting in 1986, Aharonov
 \cite{aharonov-nyas}   mentioned
a surprising new effect which combines the inequivalence of conditions
to initial conditions together with what he calls "weak' measurements.
If we set both initial and final conditions, and at the intermediate time
perform a particular type of inexact measurement of a quantity, the outcome
of that measurement can be very counter intuitive. Although the measuring
apparatus and the interaction are designed so that if the initial state
is an eigen state of the measured quantity, the outcome will be approximately
given by that value for the measured quantity, in this pre and post conditioned
experiment, the expected value for the measurement is impossible according
to all the usual tenants of quantum mechanics.

Let me make this clear by an example. Our measuring apparatus is a trivial
infinite mass free particle. it is coupled to a spin $ s$ particle ( in
my example $s=20$).  The coupling is of the form
\beq
H_I={\epsilon\over \sqrt(2)} (S_x+S_y) p \delta(t-t_0)
\ee
Ie, the interaction is such that if the initial state of the free particle
is $\psi(x)$, and the state of the spin is in an eigenstate of the operator
$S_{//}$ say with eigenvalue $\sigma$, then the final state of the free
particle is $\psi(x-\epsilon\sigma)$. Thus by measuring the displacement
of the
the free particle due to the interaction one can estimate $\sigma$ and
thus measure $S_{//}$. If the particle begins in an eigenstate  ( or almost
and eigenstate) of $X$ (ie $\psi$ is sharply peaked about some value $x_0$
with an uncertainty much less than $\epsilon$) then the displacement during
the interaction can be measured precisely by determining the value $x$
of $X$ after the interaction, and $\sigma=   (x-x_0)/\epsilon$ will be
a measurement of $S_{\\}$. On the other hand, if the initial $\psi $ has
a spread of $\Delta x$, then the final determination of $X$ will give the
displacement only to $\pm\Delta x$. Ie, we will have $\sigma=(x-x_0)/\epsilon
\pm \Delta x/\epsilon$.
This is the sense in which the measuring apparatus is inexact. The
determination
of some variable of the measuring apparatus only gives an inexact estimate
of the value of some dynamic variable of the system.

Now consider the following situation. Set conditions such that before the
interaction with the measuring apparatus, the value of $S_x$ is known to
be its maximum possible value, $s$. Furthermore after the interaction,
the value of the y component, $S_y$, is known to be the maximum possible
value, $s$. What is the distribution of possible outcomes for the measuring
apparatus? One would expect this to be something like the some probability
distribution over the various possible values for $ S_{//}$ convoluted
with the initial probability distribution for the position of the free
particle.
Ie, one would expect something like $\sum_\sigma P_\sigma
|\psi(x-\epsilon\sigma)|^2$
where $P_\sigma$ is a probability for the spin to have value $-s\le \sigma\le
s$.
In particular, the average value ( expectation value) for $X$ should lie
somewhere between $x_0-\epsilon s$ and $x_0+\epsilon s$. If the measurement
is sufficiently accurate this expectation is fulfilled. Figure 1 plots
the probability distribution for the location of the particle ( $x_0=0$
and $\epsilon=1$) in the case where the initial spread of the wave function
for the particle is small. However, Figure 2 is the plot of the distribution
for the value of the position of the particle in the case where the initial
spread for $X$ is large (of order $\pm \epsilon \sqrt(s)$). Note that the
center of the probability distribution is at $x\approx 28$, and the probability
that $x$ would lie between -20  and 20, the naively expected values, is
very small. Using the determined value of $X$ to infer the value of $S_{//}$
gives a value at all times larger than the maximum eigenvalue of $S_{//}$.

\figloc1{The probability distribution for the pointer of the measuring
apparatus with pre and post conditions with a small error ($\approx .5$)
for the measurement of the spin. The maximum value of the spin is 20.}

\figloc2{The probablility distribution for the pointer with the same conditions
as in figure 1 but with a large error (5) for the infered value of the spin.
Note that the distribution centers around the value of 28 and has only a very
small probability of lying between 20 and -20.}

Note that if we regard $S$ as a classical vector spin, and we know that
$S_x $ and $S_y$ both have value $s$, then $S_{//}$ will have value
$\sqrt{S_x^2+S_y^2}=\sqrt{2} s\approx 28$.
Note also that this works in this way only if the initial state $\psi(x)$
is sufficiently smooth. (In my case I have chosen it to be a gaussian).
In particular, sharp features in $\psi$ will destroy this property.

One reaction to this example is that it is not a real measurement. However,
it meets all of the criteria of a measuring apparatus, in that if the state
of the spin is an eigenstate, the measuring apparatus produces the value
to the accuracy to which the apparatus is designed. What we have here is
a strange result which arises from the combination of an inexact measuring
apparatus, combined with the inequivalence of conditions in quantum mechanics
to be equivalent to initial conditions. (For any initial conditions, the
expectations that the result would simply have been the sum of the
probabilities
of the result for the eigenstates would have been true.) Note that this
is a measurement situation in which the measurement is not equivalent to
any determination on the spin system itself.

There is another measurement situation which leads to results in conflict
with the von Neumann equivalence of measurements and determination. This
is a situation I call adiabatic measurement. It arises out another situation
noted by the group around Aharonov\cite{AAV}. (They call it `protected'
measurements,
a term I feel to be highly misleading. They furthermore use it to argue
that the wave function is `real' in some sense, a conclusion I also have
great difficulty with\cite{unruh94}.)  This is a situation in which the
measuring apparatus
is coupled to the system sufficiently weakly, and the system's evolution
during the interaction with the measuring apparatus is dominated by a
Hamiltonian
with sufficiently widely spaced energy levels that the interaction with
the apparatus can be treated throughout as an adiabatic perturbation.

Consider a system whose Hamiltonian during the course of the interaction
is
given by $H_0.$ Consider couplings to a set of measuring apparatuses (
which for simplicity we will take as free infinitely massive particles
again, although nothing changes if we use more complex measuring apparatuses).
\beq
H=H_0 + \sum_i\epsilon_i(t) A_i P_i
\ee
where the $A_i$ are a variety of operators associated with the system (in
general non-commuting) and the $p_i$ are the momenta of a set of free
infinitely
massive particles.  We can solve this assuming that the measuring apparatus
are in the momentum eigenstates $|p_i>$, to obtain the adiabatic approximation
to the Schroedinger equation for the system
\ba
|\psi(t)>= \sum_E a_E(t)|E>
\\
i\dot a_E = E(t) a_E(t) - i\sum_{E'}a_{E'}<E(t)|\dot{|E'(t)>}\approx E(t)
a_{E}
\ea
where
\beq
E(t) \approx E_0 + \sum_i \epsilon(t) <E_0|A_i|E_0> p_i
\ee
where $|E_0>$ are the eigenstates of $H_0$ and $E_0$ their eigenvalue.
Thus the equation of motion for the state of the system plus measuring
apparatus can be written as

\ba
|\Psi(t)> = \sum_E a_E(0) e^{i\int^t (E_0 +\sum_i\epsilon(t)<E_0|A_i|E_0>
P_i dt) } |E(t)>\prod_i|\phi_i> \\
=\sum_E e^{i E_0 t} |E(t)> \prod_i e^{i \int \epsilon(t) dt <E_0|A_i|E_0>
P_i}\phi_i(x_i)\\
=\sum_E e^{i E_0 t} |E(t)> \prod_i  \phi_i(x_i- \int\epsilon_i(t) dt
<E_0|A_i|E_0>)
\ea

After the interaction with the apparatus is finished, the state is
\beq
|\Psi(t)> = \sum_{E_0} a_E(0)e^{i E_0 t} |E_0> \prod_i |\phi(x_i-\int\epsilon_i
dt <E_0|A_i|E_0>)>
\ee
Each of the measuring apparatuses has been displaced by an amount
$\int\epsilon_i(t)dt <E_0|A_i|E_0>$,
ie by an amount proportional to the expectation value of the measured operator
$A_i$  in the state $|E_0>$. Now, if we assume that the states $\phi(x_i)$
are sufficiently narrow that there is at least one $A_i$ such that $\int
\epsilon_i(t)dt (<E_0|A_i|E_0>-<E'_0|A_i|E'_0>$
is larger than the initial uncertainty in $\phi_i(x)$, then the various
energy eigenvalues will decohere. The measuring apparatuses will point
to a value
$<E_0|A_i|E_0>$, ie an expectation value, for some value of $E_0$, with the
probability of that $E_0$ given by $|a_i(0)|^2=
|<E_0|\psi>|^2$.

There are a number of strange features of this result. In the first place,
the value to which the measuring apparatus points is not that corresponding
to one of the eigenvalues of $A_i$. The measuring apparatus measures $A_i$,
but the pointer does not give one of $A_i$'s eigenvalues but rather gives
an expectation value, $<E_0|A_i|E_0>$ in any single measurement. Furthermore
if we repeat the experiment, we will, as expected get a variety of answers
that the pointer points to , namely each of the various expectation values
for the various possible values of $E_0$. Over a large number of trials,
we expect to get a number of trials in which we get a specific value
$<E_0|A_i|E_0>$
a number of times given by $N|<E_0|\psi>|^2$ times. Thus the statistical
expectation value for the measurements of $A_i$ are
\beq
<A_i>_{stat} = \sum_{E_0} |<psi|E_0><E_0|A_i|E_0><E_0|\psi>
\ee
But the quantum mechanical expectation value of $A_i$ is given by
\beq
<A_i>_{QM} \sum_{E_0}\sum_{E'_0}|<psi|E_0><E_0|A_i|E'_0><E'_0|\psi>
\ee
In general, only if the vectors $E_0$ are also eigenvectors of $A_i$ are
these two expressions the same. I.e., the statistical expectation value
of $A_i$ obtained by performing a large number of such adiabatic measurements
is {\bf not} the quantum expectation of $A_i$ in the state of the system.

We thus have a situation which violates almost all of the standard lore
about measurements. Since the $A_i$ are not necessarily commuting ( there
is nothing in the above derivation which demands that they commute), we
can, in a single measurement measure non-commuting variables. Furthermore,
if the initial state is an eigenstate of $H_0$, then every measurement in
and ensemble of measurements will give exactly the same value for the
measurement
of those non-commuting variables. there will be no statistical uncertainty
in the result.  Furthermore, the outcome of the measurement is {\bf not}
an eigenvalue of the operator corresponding to the measured quantity $A_i$.
It is rather an expectation value of that quantity. The statistical
distribution
of the results does not depend on the quantities $A_i$ being measured.
Rather, the statistical distribution depends on the eigenvectors of the
Hamiltonian $H_0$ which is {\bf not} coupled to any measuring apparatus
at all.

It is interesting to note that the standard von Neumann measurement falls
into exactly this class as well. In the von Neumann measurement, the
interaction
with the measuring apparatus is such that the coupling to the apparatus
dominates the dynamics during the measurement. eg, the hamiltonian is of
the form
\beq
H= H_{free} +\epsilon \delta(t) A P
\ee
In this case, the dominant hamiltonian during the interaction is $A$, since
$\delta(0)$ is infinite. The coupling to the measuring apparatus $A$ clearly
commutes with the dominant Hamiltonian $A$ and thus the interaction is
adiabatic for an arbitrary time dependence of $\epsilon(t)=\epsilon \delta(t)$.
According to our adiabatic analysis, the measurement will give us various
expectation values $<E_0|A|E_0>$ where the $E_0$ are the eigenvalues of
the dominant hamiltonian $A$. I.e., the $E_0$ are just the eigenvalues
$a$ of $A$. Thus the measured quantities will be $<a|A|a> =a$ the eigenvalues
of $A$. The probability of obtaining the value of $a$ in the measurement
is $|<E_0|\psi>|^2=|<a|\psi>|^2$, and the statistical expectation value
of $A$ is
\beq
<A>_{stat}= |<\psi|a>|^2 a = <A>_{QM}
\ee
We thus see that the usual rules on measurement are simply a special case
of the results obtained for adiabatic measurements.

Note however that the general adiabatic measurement is not equivalent to
a determination. This however does not make them any the less interesting
as measurements. In fact the archetypal quantum measurement example, the
Stern Gerlach experiment, used in almost all the text books as an example
of the von Neumann measurement is actually an adiabatic measurement, in
which non-commuting observables, the spin in both of the transverse directions
is adiabatically measured. For details see reference \cite{unruh94}.

\section{Conclusions}

The key points of this talk have been

1) in the standard formulation of quantum mechanics the term measurement
is used to denote two distinct concepts. In order to clarify the problems,
I have suggested that it would be useful to use separate terms to denote
separate concepts, and have proposed that we use `determination' for the
axiomatic concept and reserve measurement for the physical notion of using
changes induced into one system to deduce properties of another system.

2) I have pointed out the old but little understood feature of quantum
mechanics
that the conditions in quantum mechanics are not equivalent to initial
conditions. A couple of examples have emphasized this unexpected nature
of the results obtained in quantum mechanics when conditions span the the
time during which one wants to ask questions of the quantum system.

3) I have shown that if we liberate the notion of measurement from
determination,
the variety of measurements are in fact much larger than simply those which
are equivalent to a determination. although this has been well known for
a long time in the case of inexact measurements, the example of adiabatic
measurements show that many of the features of measurements of the von
Neumann type are features restricted to that type of measurement alone.

\acknowledgments
I would like to thank Y. Aharonov and L Vaidman for discussion. I would
also thank the Canadian Institute for Advanced Research for a fellowship
and other support during the course of this work. This work was also performed
under an grant from the Natural Science and Engineering Research Council
of Canada.

\end{document}